\documentclass[5p]{elsarticle} 

\usepackage{hyperref} 
\usepackage{amssymb}
\usepackage{amsmath}
\usepackage{array}

\begin{document}

\begin{frontmatter}

\title{A large area high resolution imaging detector for fast atom diffraction.}

\author{Sylvain Lupone, Pierre Soulisse and Philippe Roncin}
\address{Institut des Sciences Mol\'{e}culaires d'Orsay (ISMO), CNRS, Univ. Paris-Sud, Universit\'{e} Paris-Saclay, F-91405 Orsay, France}

\begin{abstract}
We describe a high resolution imaging detector based on a single 80 mm micro-channel-plate (MCP) and a phosphor screen mounted on a UHV flange of only 100 mm inner diameter. 
It relies on standard components and we describe its performance with one or two MCPs.
A resolution of 80 $\mu$m rms is observed on the beam profile. 
At low count rate, individual impact can be pinpointed with few $\mu$m accuracy but the resolution is probably limited by the MCP channel diameter. 
The detector has been used to record the diffraction of fast atoms at grazing incidence on crystal surfaces (GIFAD), a technique probing the electronic density of the topmost layer only. 
The detector was also used to record the scattering profile during azimuthal scan of the crystal to produce triangulation curves revealing the surface crystallographic directions of molecular layers.
It should also be compatible with reflection high energy electron (RHEED) experiment when fragile surfaces require a low exposure to the electron beam. 
The discussions on the mode of operation specific to diffraction experiments apply also to commercial detectors.

\end{abstract}

\begin{keyword}
fast atom diffraction, reflexion high energy electron diffraction, imaging detector.
\end{keyword}

\end{frontmatter}

\section{Introduction}
\begin{figure}	\includegraphics[width=0.7\linewidth]{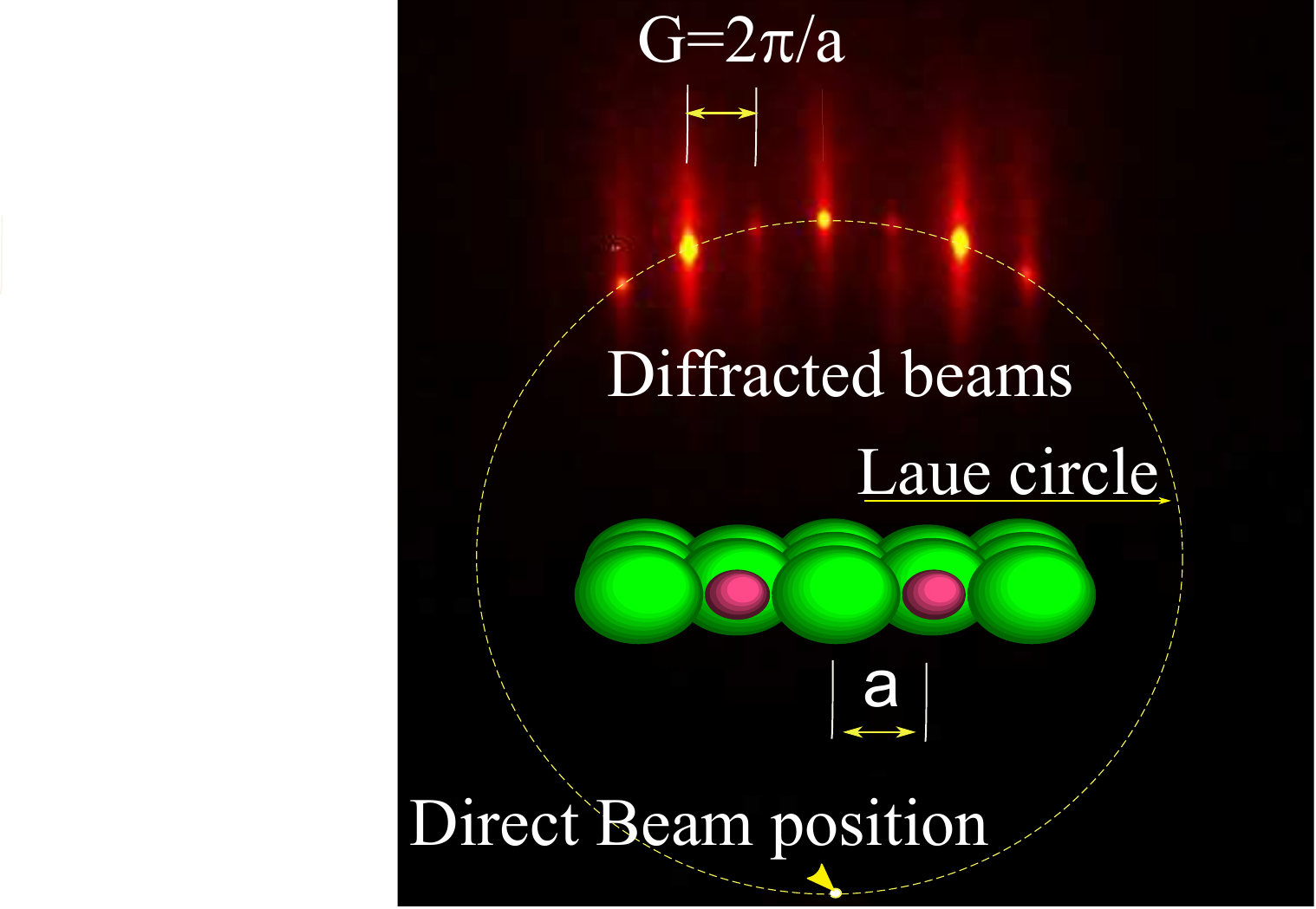}
	\caption{Diffracted beams of 500 eV He on a LiF surface \cite{Debiossac_PRL} recorded with the present detector. The bright spot on the Laue circle correspond to elastic diffraction and the extension to inelastic diffraction \cite{Roncin_PRB_2017}. The intensity can be concentrated on few, very bright spots but the exact value of the faint spots is equally important.
		\label{fig_LiF}}
\end{figure}

Grazing incidence fast atom diffraction uses light atomic or molecular projectiles in the keV energy range to probe crystalline surfaces with the same geometry as RHEED.
However, since keV helium atoms do not produce significant amount of light when hitting a phosphor screen, a specific detector is needed to image the impacts (Fig.\ref{fig_LiF}).
So far only micro-channel plates, originally developed for high performance night vision systems, can efficiently detect keV atoms by producing an avalanche of few $10^3$ electrons at the output of the impacted channel.
Several options are available to image this electron shower together with the associated arrival time useful for time of flight measurements and/or for coincidence experiment (see e.g. the extensive work by J.S. Lapington and O.Jagutzki \cite{Jagutzki_2002}).
\begin{figure}	\includegraphics[width=0.95\linewidth]{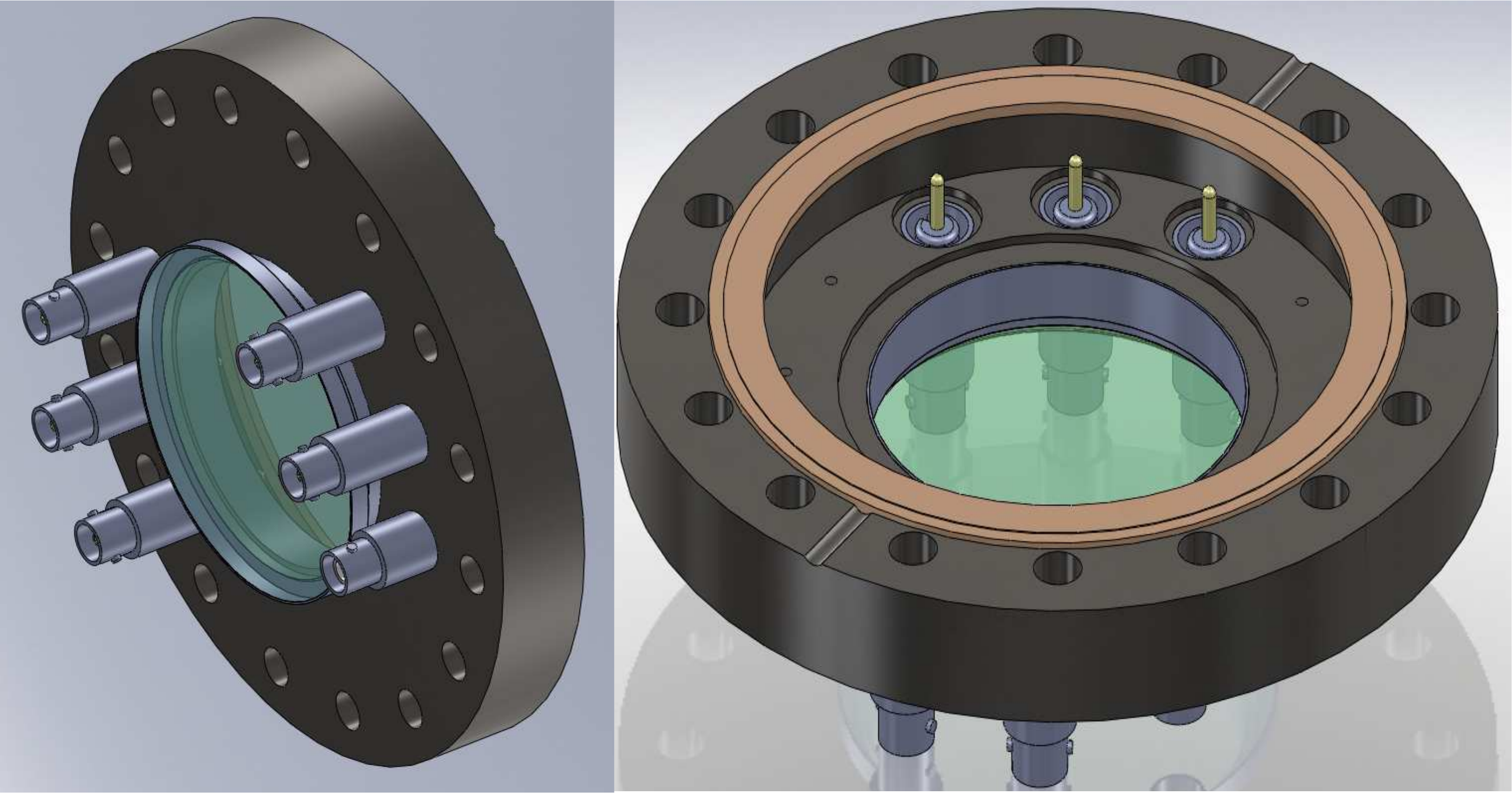}
	\caption{Shematic view of the air side (left) and vacuum side (right) of the DN100CF UHV flange supporting the detector.
		\label{fig_bride}}
\end{figure} 
One of the most universal is probably the delay-line double MCP\cite{Roentdek} which have been used by the Berlin group\cite{Winter_PSS_2011} since a decade to record scattering profiles including diffraction experiments. 
However, these detectors are comparatively more complex than simple imaging detectors in particular significant local distortions are often present\cite{Jagutzki_2002,Flechard_97}. 
These local aberrations can be empirically corrected to an impressive absolute positioning accuracy of 10 $\mu$m rms \cite{Flechard_2016} but with additional efforts including a permanent specific grid. 
When neither timing accuracy nor coincidence measurements are needed, a simpler technique almost aberration free, is to project the electron shower onto a phosphor screen and to image it by a camera. 
Such detectors are commercially available\cite{proxivision} with optimal performances (using a larger 150 CF flange); the count rate can be very high and the ultimate resolution is usually limited by the MCP hole spacing. 
For GIFAD experiments inside a molecular beam epitaxy vessel, where the environment can be harsh and where proven solutions are mandatory, such a detector gave full satisfaction\cite{Debiossac_PRB_2014,Atkinson_2014,Debiossac_ASS_2017}. 

For experiments where the possibility of breaking the vacuum is not a dramatic issue or where the space is limited, a more compact home-made version using standard MCP and phosphor screen is described allowing easy maintenance or replacement of MCP and phosphor screen and no specific HV connectors.
It is based on our experience with imaging MCP detectors, with capacitive division \cite{Roncin_86,Roncin_1987}, resistive division  \cite{Rousseau_2008,Lupone_2015}, geometric division \cite{Flechard_97}, delay-line\cite{picard_2005,Cassimi_2009}, muti-detector array\cite{Morosov_96,Roncin2002} and fast phosphor \cite{Urbain_2015}. 
This detector has been machined in our workshop and has been used for GIFAD experiments \cite{Debiossac_PRL,Zugarramurdi_2013,Debiossac_2016}. 
It is described below together with a discussion on operating conditions using either one or two MCP and a CCD or a CMOS camera. 
The same detector can be used to for recently developed atomic triangulation techniques\cite{Nataliya,Feiten_2015} where the fwhm of the scattering profile of grazing incidence atoms is recorded as a function of the surface azimuthal angle to reveal the crystallographic directions where molecules are aligned. 
It can be useful in RHEED setup where a low electron flux is desired to limit the damage to the surface induced by the electron beam.

\section{detector design} 
The detection of keV atoms requires at least one micro-channel plate to convert single particle into an "electron shower" of few thousands electrons emitted at the output of the hit channel. 
The ultimate resolution is therefore limited by the channel spacing that can be chosen between 10 and 20 $\mu$m. 
To get close to this ultimate resolution in standard accumulating mode, one must limit the coulombic blow-up of the electron shower, assumed to leave the channel within a cone with a 45 degree angle. 
The recommended proximity focusing consist in a acceleration this cone by two to four kilovolts toward the phosphor screen placed at a distance less than a mm from the channel plate. 
This is the delicate aspect of the assembly. 

\subsection{mechanical design}
\begin{figure}	\includegraphics[width=0.9\linewidth]{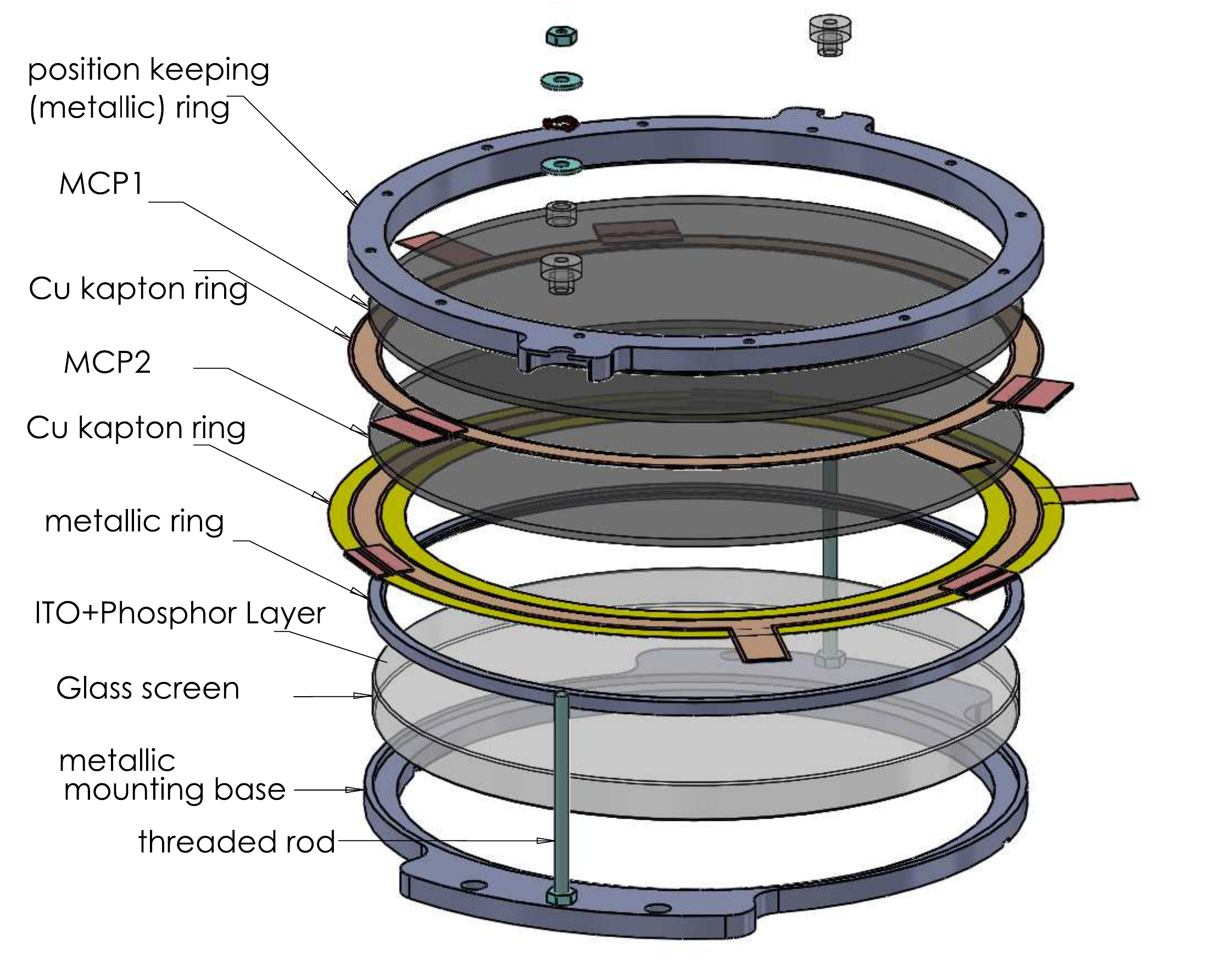}
	\caption{schematic view of the detector assembly with two MCP and a phosphor screen. The detector is assembled separately and fitted inside the UHV flange in Fig.\ref{fig_bride}. For operation above 5 kV, the two threaded screws can be replaced by ceramic screws.
		\label{fig_MCP}}
\end{figure}

The design is based on a blank UHV flange with inner diameter of 100 mm (6" O.D.), a 68.4 mm inner diameter weldable glass view-port\cite{viewport}, a 80mm phosphor screen\cite{phosphor} and six standard high voltage MHV or SHV weldable feed-through. Fig.\ref{fig_bride} displays drawings of the UHV flange with the view-port shifted outside to leave room for the detector assembly. 
The phosphor screen is very fragile and should never be touched. 
For this reason the detector itself, holding both the MCPs and the phosphor screen has been designed to be assembled separately before mounting on the flange. 
It is sketched in Fig.\ref{fig_MCP} and provides safe handling of the expensive parts ; MCPs and screen.

Note also that the UHV flange holding the view-port and connectors is also compatible with 50 mm MCPs and phosphor screen offering a significantly cheaper option while maintaining a limited overall detector dimension. 

\subsection{electrical design} 
The bias of the different parts is achieve with kapton film covered with a 15 $\mu$m Cu tracks. 
These are easily designed by drawing software and, at least in single sided deposit, these can be purchased directly from the electronic printed circuit board industry including the useful precision laser cut. 
The distance between MCP and screen is well controlled by these kapton rings which also provide very good parallelism. 
This nice and simple feature is also the major limitation of the assembly as the electrical path between MCP and screen limits the electron acceleration voltage onto the screen to 3 kV.

We have also tried to use a hat shape design (see fig.\ref{fig_hat}) where the screen is on top and where the bias voltage is applied on the base, one or two mm below allowing a much larger electrical path between MCP and screen and therefore higher voltage. 
Unfortunately we did not find standard supply and our custom order with an ITO layer did not show correct electrical continuity.
The assembly is compatible both with flat disk or hat shape design of the phosphor screen, but the later requires an additional specific glass or ceramic spacer always difficult to order in small quantities. 

\subsection{phosphor screen and imaging system} 
The phosphor screen is a standard 3-4 mm thick cylindrical disk coated with an ITO layer and a 4 $\mu$m P43 Phosphor layer \cite{phosphor} that have a high efficiency of less than 20 eV of electron energy to emit a photon. 
These are easily available with a high uniformity and a high spatial resolution.
The camera lens is a high aperture  F-0.95, 17 mm focal length manual lens.
It is in charge of collecting as many of the emitted photons as possible while preserving the spatial information.
The second requirement being more important, the camera is place at 250 mm from the screen giving a maximum angle of $\sim$ 11 deg on the lens to reduce optical aberrations. 
As a result the solid angle or geometric collection efficiency of only 5 $10^{-4}$. 
It is relatively easy to zoom on a particular part of the screen but this requires a time consuming calibration. 
With modern camera equipped with sensors having more than 1000x1000 pixels, optical zooming was not needed.
The quantum efficiency for the 500-600 nm photons emitted by the P43 screen is excellent, above 70\%. On the other hand, the dynamic range as well as the intrinsic noise are still limiting factors that will be discussed below.

\subsection{power supplies} 

To allow full flexibility to favor detection of positive or negative ions particle or to repel them, the front voltage is not grounded from the inside. The biasing of one or two MCP is achieved with a budget miniature 5 W high voltage DC/HV 12 V/3 kV power modules which output can be floated by $\pm$ 3 kV with respect to the ground. 
The connections to the MCP incorporates a 1 M$\Omega$ resistors in series limiting the current through the MCP in case of a vacuum failure.
More specific, to avoid possible damage to the phosphor screen during voltage breakdown, the screen is biased by a power supply with a 100 $\mu$A trip current limit.
In principle it could be replaced by a 40 M$\Omega$ resistor in series but we have not tried this option.


\subsection{efficiency and saturation with one and two MCPs}

For one MCP, the number of emitted electrons per detected impact is on the order of 5 $10^3$. 
With a 2 kV acceleration to the P43 phosphor screen, each electron produces a hundred photons emitted in all directions. 
Among these 5 $10^5$ photons, only few hundreds reach the CCD or CMOS sensor. 
For two MCP, the electron gain is around $10^7$ giving few $10^5$ photons on the sensor which is extremely comfortable in terms of sensitivity but more restrictive in terms of single image dynamics because spatio-temporal overlap of the electron cascade rapidly affect the second MCP.
The dead time of a single MCP channel is estimated around few $10^{-2}$ s \cite{wiza} limiting the rate to 100 Hz per individual pore in single MCP operation.
Assuming a 200 $\mu$m geometrical size of the diffraction spot and 20 $~ m$ MCP pore size the maximum rate for linear behavior is therefore on the order of 10 kHz per spot. For a 2 MCP chevron assembly, the inter MCP assembly is usually optimized for high gain so that the electron shower of the first MCP hits the second MCP over a thousand of channels i.e. over a spot of $\sim$ 100 $\mu$m therefore limiting the linear regime below a kHz. The situation can be improved by reducing the gain but checking the linear behavior of the two MCP assembly is certainly a good practice. 
The performance of one and two MCP will be discussed below in connection with the camera and software.


\section{Two MCPs operation}
By fear of a too low light level, the detector was first equipped with two MCPs providing a very high contrast of each impact and allowing rather long integration time of 10 s while preserving the ability to isolate a single count from the background as illustrated in Fig. \ref*{fig_Dark} a). 
For such long integration time, the seven year old Hamamatsu C8484 CCD camera shows some scatter hot spots and a non uniform background noise level.

In order to test the detector with real impacts we first generate ions by opening slightly a manual gate valve on top of an ion-pump located one meter away.

The scattered impacts in Fig. \ref*{fig_Dark}a are well fitted by 2D gaussian profile having a wide dispersion in amplitude but a narrow distribution in width around $\sigma$ = 1.2 $\pm$ 0.2 pixel rms wich corresponds to an angular resolution of 0.0055 deg rms and a spatial extend of 85 $\mu$m rms (195 $\mu$m fwhm).

As long as the density of individual spots is low enough to avoid overlap, it is relatively easy to pinpoint each impact and to fit its position with a resolution of $\sigma /\sqrt{N}$ providing a theoretical value close to a $\mu$m.
In practice the resolution is certainly limited by the pore size but this remains to be demonstrated by capturing a large enough number of successive images to achieve a density of few impacts per $\mu$m$^2$. 
We did not invest efforts in this direction because, so far, our resolution is limited by the primary beam size.
For diffraction application it is expressed in angular divergence and, for neutrals the only way to limit this angular spread is by using a set of two diaphragms.
Schematically, the first one $\phi_1$ defines a source size and the other one $\phi_2$, $L=$50 cm apart, defines the angular acceptance $\delta \theta =(\phi_1 + \phi_2)/L$. 
Using identical diameter $\phi$ the transmitted intensity drops drastically with $\phi^4$ ($\phi^2$ for each diaphragm) so improving the resolution by a factor two require a intensity reduction by a factor 16.

Two sets of identical diaphragms were used $\phi$=100 $\mu$m and $\phi$=50 $\mu$m and the measured standard deviation of the primary beam spot profile were  $\sigma=11.5$ mdeg and $\sigma=7.5$ mdeg respectively. 
Interpreting this angular resolution as a combination of the spatial resolution and of the angular divergence adding up quadratically, an angular divergence of 9.6 mdeg and 5 mdeg is derived for $\phi$=100 $\mu$m and 50 $\mu$m respectively, in good agreement with the simple estimate $\theta_{res}\sim \phi/L$.
Consistently, the spatial resolution is estimated at of $\sigma=100 \pm 10$ $\mu$m rms partly limited by the CCD pixel corresponding here to 70 $\mu$m on the phosphor screen. 
In term of angular resolution it corresponds to $\sigma=6.8$ mdeg which is the limiting value for simple operation without any treatment.

Due to the limited 12 bits dynamic range of the camera \cite{high dynamic}, pixel saturation can take place rapidly. 
This is easily identified by the characteristic plateau at the maximum pixel capacity alerting that a saturation is present.
One has to pay attention that the plateau may not be visible on compressed image or on projected profiles because the amount of saturated pixel is less so that the clear plateau at the pixel level becomes a smooth flattening of the peak.
This saturation can be bypassed by the internal software of the camera allowing successive shortly exposed images to be added before transmission but this have to be adapted carefully to the actual count rate. 
It could appear safe to have a very short exposure time avoiding the saturation to occur on any individual image but short exposure also means faster readout and higher electronic noise of the analog to digital converter that accumulates in the integrated image. 
This can usually be neglected with cooled camera and 2 MCP operation. 
The worst case situation corresponds to an integrated sequence where the single image exposure time was chosen too close to the saturation limit of the brightest spot. 
Only some images will be saturated so that the integrated image will not display any obvious sign of saturation while its intensity is obviously affected.
This may happen when the intensity ratio of a bright spot saturating rapidly and a very faint spot have to be recorded accurately. 
In addition, pixel saturation is often associated with the blooming effect where part of the excess electrons from a saturated pixel flow to adjacent pixels.
In general, the lost intensity information can not be easily recovered.

\begin{figure}	\includegraphics[width=0.95\linewidth]{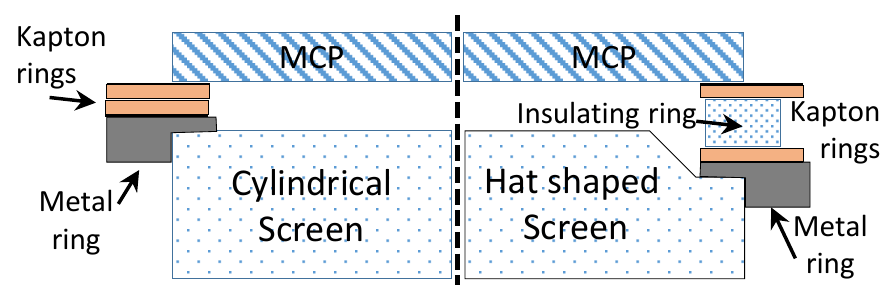}
	\caption{Schematic view of the cylindrical screen used here (left side) and tho hat shape screen (right side). The later allows a larger distance to bias the output of the MCP to the screen by few kV.
		\label{fig_hat}}
\end{figure}

\section{Single MCP operation}
The Fig.\ref{fig_Dark}b) shows an image of the ions from the ion pump recorded in 250 ms by a Hamamatsu CMOS ORCA-Flash 2.8 camera. 
The individual impacts remain clearly visible but since the spot intensity is much lower, the less intense spots start to fade away, embedded in the background noise after few 10 seconds of integration.
The process is stochastic and increasing the camera gain or adding up successive images does not solve the problem. 

The individual spot size is measured around 70 $\mu$m rms slightly less than with two MCPs but since the intensity is less the single spot pointing ability is less.
The most attractive aspect of single MCP operation is probably the high rate behavior is expected to be much better because the probability that the same MCP channel is hit twice during the dead-time of $10^{-2}$ s is now 100 to 1000 times less than with two MCP. 

For instance comparing the beam intensity with the single particle response we have estimated a beam intensity of  900$\pm$100 counts per second though the 50 $\mu$m diaphragms. The measured beam profile of 85 $\mu$m rms indicates a surface of 0.03 mm$^2$ corresponding to a very large particle density of few $10^6$ particles per second per cm$^2$.
\subsection{single particle mode}
Ideally, a single particle counting software should be able to identify spots on top of a not uniform background and to extract its main properties;  intensity widths and precise location.
Such software is not yet available inside the camera \cite{Post_treatment} but the situation could evolve rapidly with the development of single-particle tracking and super-resolution localization microscopy \cite{Bourg_2015}.

\section{dynamic range}
In diffraction measurements, the form factor revealing the information on the unit cell content is hidden in the relative peak intensities so that the ability to measure as accurately as possible these peak intensities is a very important characteristic. 
In GIFAD, the problem is quite serious because only one Laue circle is usually observed \cite{Debiossac_PRL} and the spot size is usually on the order of few hundred microns as on Fig.\ref{fig_LiF}. 
Depending on the lattice parameter and projectile perpendicular energy, the number of diffraction peaks energetically allowed is usually limited.
So far, the maximum number of well resolved diffraction peaks did not exceed a hundred\cite{Debiossac_PRB_2014}. 
As a result, the intensity is highly concentrated so that various saturation effect can produce non linearities degrading the dynamic range. 

\section{Limitations} 
The two MCP setup is close to the saturation level of the second MCP when the experiment produces only two to three diffraction spots \cite{Debiossac_PRL}. These saturations effects are independent of the frame rate since they occur before the phosphor screen. They affect the accuracy of the intensity measurements and can easily go unnoticed.
Camera saturation can also occur easily but are usually rapidly detected by the flat top of the peaks or by the camera software.

For applications where a high flux is expected on a dilute scattering pattern, for instance in triangulation technique, the resolution requirement are less severe because the scattering profile is usually broader. 
The spot density is then much lower so that saturation effects are less likely to occur.

\subsection{CCD and CMOS camera}
\begin{figure}	\includegraphics[width=0.99\linewidth]{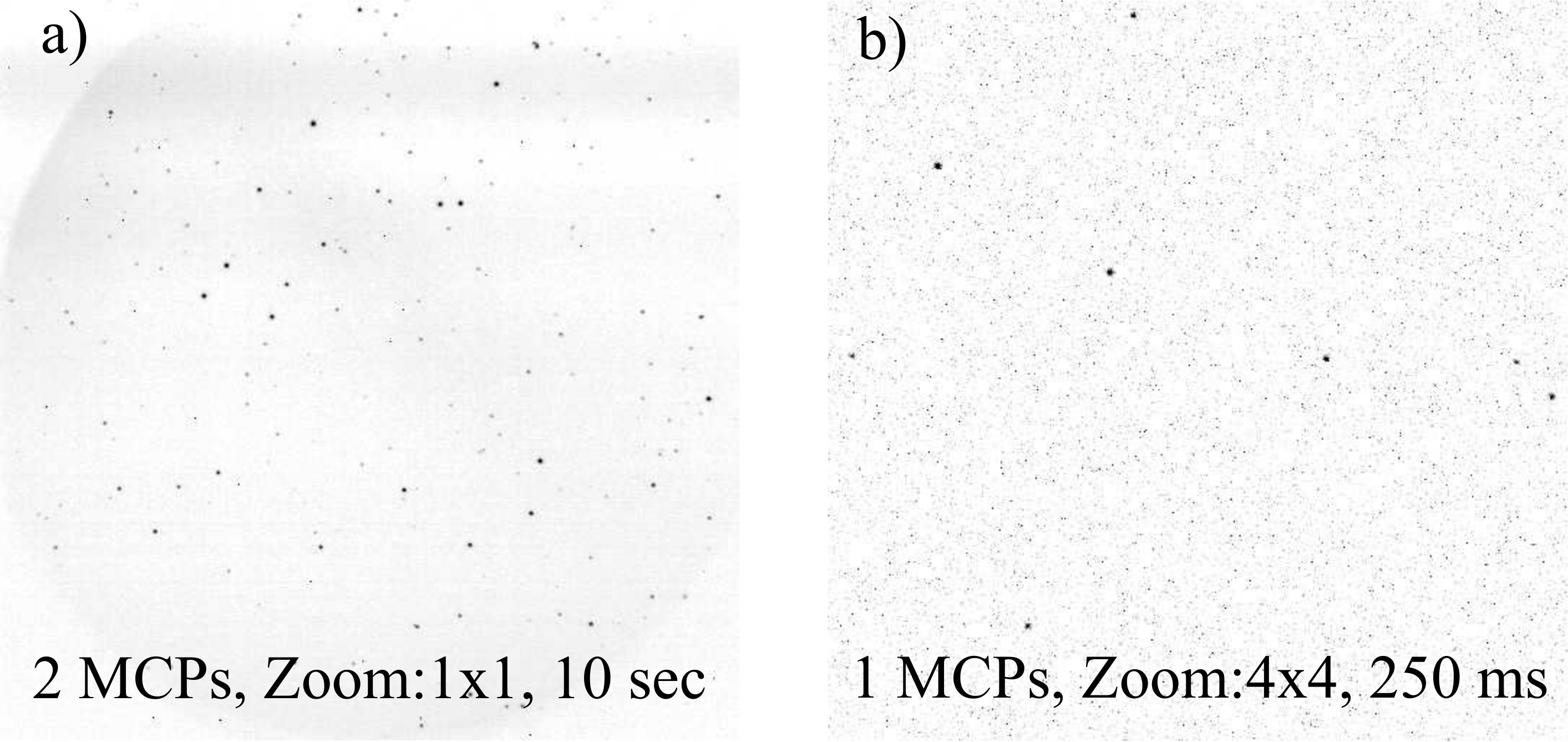}
	\caption{image of the dark counts registered with one or two MCPs. With two MCPs (left) the individual spots are broader ($\sim$90 $\mu$m rms) but remain visible even after 10 sec integration. With only one MCP (right) the spots are smaller ($\sim$ 65 $\mu$m rms) but with less intensity so that a shorter integration is required to keep the spot intensity above the fluctuations of the background level.
		\label{fig_Dark}}
\end{figure}

Ideally the imaging system should be transparent bringing a natural interface to the computer without limitation.
The performance of CCD and CMOS camera have made significant progress with very high detection efficiency for green photons but the dynamic range is still limited to 12 bits so that care should be taken to adapt the camera operation to the MCP assembly. 
If the $10^5$ photons collected from the two MPC assembly are focused on a single pixel, one impact is enough to reach saturation limiting the associated count rate to one impact per frame.
In Fig.\ref{fig_Dark} our older CCD camera is perfectly adapted to a double MCP operation and reference image subtraction.
Ideally, after each day of operation reference images are recorded with the same exposure time as used during measurements.
These images are then automatically subtracted from the diffraction images before analysis.
We kept the procedure with our recent CMOS camera but the Fig.\ref{fig_Dark}b) shows a rather flat noise. Neither Fig.\ref{fig_Dark}a) nor Fig.\ref{fig_Dark}b) have been processed.

Last but not least, to hold a resolution of 100 $\mu$m over 80mm require a CCD camera with a minimum of 800x800 pixel with one pixel to describe a spot. In practice a 2000x2000 is  obviously a better choice.

\subsection{signal to noise ratio}
In this section the S/N discussion is restricted to the ability of identifying single impact.
The signal intensity is mainly dependent on the number of MCP, the phosphor screen acceleration and the collection efficiency with a very small contribution to the noise.
The camera itself has two different sources of noise. 
The readout noise is usually low enough, unless very fast readout is required, but the steady noise per unit time will eventually accumulate above the amount of light produced by a single impact.
In general, it is important to consider the system as a whole including the software limitations as well as the analyzing procedures \cite{Debiossac_Nim_2016}.

The low level counting ability has been investigated by observing background counts from ion emitted by an ion pump far from the detector.
The rate of random individual ion impacts was adjusted by a manual gate valve on top of the pump. 
The Fig.\ref{fig_Dark} shows such low count rate images for one and two MCPs.
With two MCPs the amount of light is so high that single impact remain visible even after integration time of minutes.

When only one MCP is used the number of photon collected per impact is reduced to $\sim$ hundred and if it is spread over few pixels, the associated level can disappear below the noise level in few second of minutes depending on the background level. This is the opposite limitation than the rapid saturation of a two MCP assembly. It can be circumvented by using the single particle software yielding quasi infinite dynamic range.
more precisely, it is not the exact level of the background noise but its fluctuation that will progressively blur the individual impact.

\subsection{Background light}
Finally, it should be mentioned that the low acceleration voltage to the phosphor screen used here pushed us to select a screen without the optional thin Aluminum over-layer. 
Indeed, its thickness implies that electrons loose almost one keV of kinetic energy when passing through the Al layer.
For 2 kV acceleration the number of photons is divided by two and this is compensated by the factor of two provided by the mirror effect reflecting half of the photons towards the camera.
A thinner layer is possible but is more vulnerable to de-lamination. 
We underestimated the problem of stray light passing through the single MCP and illuminating the screen and therefore captured by the camera. 
This is not a problem at room temperature where shielding the view-port is enough to suppress stray light but this turned out to be a problem at high temperature because the filament heating the sample could not be sufficiently shielded. 
The amount of stray light is significantly less with two MCP but still problematic when the sample is heated to 1000$^\circ$C.

The best solution would probably be to use a single MCP detector with a hat shape phosphor screen with aluminum over-layer.

\section{summary and conclusion}
A low cost detector has been described with standard components allowing comparatively easy maintenance. 
It is compatible with standard flat disk phosphor screen and with more elaborate hat shape screen allowing higher acceleration voltages and higher luminosity (see fig.\ref{fig_hat}). 
We have measured a native resolution better than 100 $\mu$m rms. 
A specific software able to isolate and fit single spots on-line would certainly help achieving typical resolution on the order of 20 $\mu$m  providing a virtual detector with 14 Megapixels.

Depending on computational efficiency and spot density a count rate in the MHz range should be possible.

The limited amount of light produced by a single MCP requires a better camera and would benefit from a better optical collection efficiency and/or a better on-line single count software.

Double MCP assembly tend to produce too many light so that saturation effects of the camera and mainly of the second MCP are possible and require special attention.
It should also be noted that, at low count rate, the large number of emitted electrons can also be interesting for multi-hit experiments such as molecular dissociation.
Indeed, double MCP operation produces enough electrons to generate a measurable electric signal on the ITO or Al layers.   
The signal transported with a $50~\Omega$ line is then compatible with modern fast digital oscilloscopes recording small amplitudes at GHz frequencies.
Using this scheme, Urbain \textit{et al.} \cite{Urbain_2015} have been able to correlate the spot intensity measured in the camera with the pulse height measured in the scope demonstrating a high resolution time and position sensitive system with almost no dead time.
The main limitation encountered here was that 80 mm hat shape screen are not proposed in standard catalog preventing the use of acceleration voltages larger than 3 kV. 
The performance of the detector presented here compares with that of commercial detectors and the limitations for diffraction studies, discussed in detail above are also similar.
As a main difference, the detector is supported on a more compact UHV flange and allows easy customization, for instance by designing a camera supporting system from the beginning.

\section{Acknowledgment}
Y. Picard, M. Debiossac, A. Momeni and H. Khemliche are kindly acknowledged for their useful advices.


\begin{thebibliography}{5}
\bibitem{high dynamic} Note that higher dynamic range camera are currently available, these can significantly increase the safe tuning range.
\bibitem{Post_treatment} Note that open source and commercial softwares offer solutions but mainly dedicated to post acquisition treatment.
\end{thebibliography}
\end{document}